\documentclass[preprint,aps]{revtex4}
\usepackage{epsfig,graphics,color}
\usepackage{amsthm}
\usepackage{enumerate}
\usepackage{ulem}
\usepackage{epstopdf}
\usepackage[version=4]{mhchem}
\usepackage{physics}
\usepackage{siunitx}
\usepackage{comment}

\DeclareSIUnit\molar{M}

\begin{document}

\title{Building up DNA, bit by bit: \\ a simple description of chain assembly}

\author{R. Foffi}
\email{riccardo.foffi@gmail.com}

 \author{F. Sciortino}
\email{francesco.sciortino@uniroma1.it}

\affiliation{Dipartimento di Fisica \\
Sapienza Universit\`{a} di Roma, Piazzale Moro 5, I-00185 Rome, Italy}

\author{J. M. Tavares}
\email{jose.tavares@isel.pt}

\author{P. I. C. Teixeira}
\email{piteixeira@fc.ul.pt}

\affiliation{ISEL -- Instituto Superior de Engenharia de Lisboa,
Instituto Polit\'{e}cnico de Lisboa \\
Rua Conselheiro Em\'{\i}dio Navarro 1, 1959-007 Lisboa, Portugal}
\affiliation{Centro de F\'isica Te\'orica e Computacional, Faculdade de
Ci\^{e}ncias da Universidade de Lisboa,
P-1749-016 Lisboa, Portugal}

\date{29 July 2021}

\begin{abstract}
We simulate the assembly of DNA copolymers from two types of short 
duplexes, as described by the oxDNA model. We find that the statistics 
of chain lengths can be well reproduced by a simple theory that treats 
the association of particles into ideal (i.e., non-interacting) clusters 
as a reversible chemical reaction. The reaction constants can be predicted 
either from Santalucia's theory or from Wertheim's thermodynamic perturbation 
theory of association for spherical patchy particles. Our results suggest that 
theories incorporating very limited molecular detail may be useful for 
predicting the broad equilibrium features
of copolymerisation.
\end{abstract}

\maketitle

\section{Introduction}

\label{sect-intro}

The assembly of multifunctional units into linear or branched architectures is 
a key ingredient of copolymerisation. In turn, the properties of copolymers 
depend crucially on how these units are arranged, as in alternating, random or
block copolymers \cite{Hamley:1998}. Examples are manifold, and we mention 
just a few: the stacking transition of single-strand DNA \cite{ssDNA}; the 
nature of the de-mixing instabilities in both coil-coil \cite{Teixeira:2000}
and coil-rod \cite{Teixeira:2007} polymer blends undergoing polycondensation
reactions; the ability of urethane-urea elastomers to exhibit strain-induced
periodic textures \cite{bookchapter}; the self-healing nature of 
poly(methyl methacrylate)/n-butyl acrylate over a narrow range of compositions
\cite{Urban:2018}; and the association of DNA duplexes by stacking
interactions~\cite{stacking1,stacking2,stacking3}.

The actual sequence of building blocks on individual copolymer molecules is 
experimentally inaccessible and must be inferred indirectly, e.g., from the
comonomer ratios, or from details of the synthesis method employed. It would 
be most desirable to have a predictive theory for this information that might 
be used as input to theories for macroscopic properties, e.g., elastic or 
rheological. Such a theory could be readily validated by computer simulations 
of copolymerisation. From a more practical point of view, this approach would
have the added bonus of enabling the `reverse engineering' of desired polymers: 
by elucidating which properties of building blocks (e.g., size, interaction 
energies) produce which architectures, polymer synthesis could be more 
effectively directed towards specific outcomes.

In this paper  we consider one prime application of the above: the assembly 
of DNA chains from two types of monomers, each consisting of one
short nucleotide duplex. Besides its biological 
relevance, this has two advantages with respect to testing our theory: first, 
it is the simplest case of linear aggregation; second, by tuning the model 
parameters we can adjust the interaction energies between different building 
blocks and thus generate a large variety of chain architectures, allowing for 
a more thorough and detailed comparison between simulation and theoretical 
predictions. We start in 
section \ref{sect-theory} by expounding a very simple theory of linear 
aggregation that forgoes most microscopic detail and treats the bonding of 
polymerising units as reversible chemical reactions, governed by reaction 
constants. 
The reaction constants are taken either from 
Santalucia's treatment of the nearest-neighbour model for DNA
\cite{santalucia1998unified,bommarito2000thermodynamic}, or from 
Wertheim's thermodynamic perturbation theory of association for spherical 
patchy particles \cite{Wertheim:1984a,Wertheim:1984b}. Then in section 
\ref{sect-model} we describe the microscopic DNA model used in our simulations.
Results are presented in section \ref{sect-results}, and conclusions drawn in 
section \ref{sect-conclusions}.

\section{Theory}

\label{sect-theory}

\subsection{A minimal description of linear aggregation}

\label{subsect-linear}

Our system consists of a binary mixture of $N_A$ particles of species $A$,
each decorated with two bonding sites (`patches') of type $A$, and $N_B$ 
particles of species $B$, each decorated with two bonding sites of type 
$B$, in a volume $V$. The total number of particles is thus $N=N_A+N_B$, 
and their mole fractions are $x_A\equiv N_A/N$ and $x_B\equiv N_B/N=1-x_A$.

Each of the two sites can participate in at most one bond to another site, so a
given particle (of species $A$ or $B$) can bond to at most two other particles 
(linear aggregation). We shall regard the formation of an $\alpha\beta$ bond
as a reversible chemical reaction between an unreacted  site of type $\alpha$ 
and one of type $\beta$ ($\alpha,\beta=A,B$). If we now assume that both sites 
and bonds behave as ideal gases, then the equilibrium constant for this 
reaction is given by \cite{Atkins:2014}
\begin{equation}
\label{eq:Kab}
K_{\alpha\beta}=\frac{P_{\alpha\beta}^*}{P_\alpha^*P^*_\beta}
\equiv\exp\left(-\beta\Delta G_{\alpha\beta}\right),
\end{equation}
where $P^*_k$ is the ratio of the partial pressure of sites or bonds 
of type $k$ ($k=\alpha$, $\beta$ or $\alpha\beta$) at equilibrium to some 
reference pressure $P_{ref}$, and $\Delta G_{\alpha\beta}$ is the change in Gibbs 
free energy on forming an $\alpha\beta$ bond. Because we are assuming that all 
sites and bonds $k$ behave as ideal gases, we have
\begin{eqnarray}
\label{eq:pressure1}
P^*_{\alpha\beta}&=&\frac{N_{\alpha\beta}k_BT}{P_{ref}V}, \\
\label{eq:pressure2}
P^*_{\alpha}&=&\frac{M_{\overline{\alpha}}k_BT}{P_{ref}V},
\end{eqnarray}
where $N_{\alpha\beta}$ is the number of $\alpha\beta$ bonds, $M_{\overline{\alpha}}$
is the number of unreacted sites of type $\alpha$, 
$k_B$ is Boltzmann's constant and $T$ is the temperature. 
Using equations (\ref{eq:pressure1}) and (\ref{eq:pressure2}), equation 
(\ref{eq:Kab}) can be rewritten as
\begin{equation}
\label{eq:Kab2}
K_{\alpha\beta}=\frac{N_{\alpha\beta}}{M_{\overline{\alpha}} M_{\overline{\beta}}}
\frac{P_{ref}V}{k_BT}.
\end{equation}
If we further assume that $P_{ref}$ is the pressure of the system when no 
``chemical reaction'' has occurred (i.e., when there are no aggregates but the 
same total number of sites is present) then $P_{ref} = 2N k_B T/V$ (recall there
are two sites per particle), whence
\begin{equation}
\label{eq:Kab3}
K_{\alpha\beta}=2N\frac{N_{\alpha\beta}}{M_{\overline{\alpha}} M_{\overline{\beta}}}.
\end{equation}
We note that this result can be given a microscopic interpretation: in terms of
the partition functions of species $\alpha$, $Q_\alpha$, and of $\alpha\beta$ 
dimers $Q_{\alpha\beta}$, the condition for chemical equilibrium is 
\cite{Hill:1986,RF}
\begin{equation}
	\label{eq:equilibrium}
	\frac{N_{\alpha\beta}}{M_{\overline\alpha}M_{\overline\beta}} =
	\frac{Q_{\alpha\beta}}{Q_\alpha Q_\beta}=\frac{K_{\alpha\beta}}{2N}
        =\frac{\exp\left(-\beta\Delta G_{\alpha\beta}\right)}{2N},
\end{equation}
i.e., the partition functions of particles and bonds are subsumed in the 
equilibrium constants, for which we need some prescription. We shall come 
back to this point later.

From the above we can now derive the laws of mass action for the three 
reactions: $\overline{A}+\overline{A}\rightleftharpoons AA$, 
$\overline{B}+\overline{B}\rightleftharpoons BB$ and
$\overline{A}+\overline{B}\rightleftharpoons AB$, 
where the overline denotes an unreacted site. Using the constraints 
\begin{eqnarray}
\label{eq:totA}
2N_A=2N_{AA} + N_{AB} + M_{\overline{A}}, \\
\label{eq:totB}
2N_B=2N_{BB} + N_{AB} + M_{\overline{B}}, 
\end{eqnarray}
and the usual definitions for the fractions of unbonded $A$ and $B$ sites,
\begin{equation}
\label{eq:XAXB}
X_\alpha=\frac{M_{\overline{\alpha}}}{2N_\alpha},
\end{equation}
we arrive at the following laws of mass action:
\begin{eqnarray}
\label{eq:lma01}
1-X_A&=&2x_AK_{AA}X_A^2+2x_BK_{AB}X_AX_B, \\
\label{eq:lma02}
1-X_B&=&2x_BK_{BB}X_B^2+2x_AK_{AB}X_AX_B,
\end{eqnarray}
Note that in the above expressions we are assuming that no rings are formed.
In the thermodynamic limit ($N\to\infty$, $V\to\infty$), $p_\alpha=1-X_\alpha$
is the probability that a site of type $\alpha$ has reacted (i.e., is bonded 
to another site). Noting that the total number of sites of type $\alpha$ that 
participate in $\alpha\beta$ bonds is $(1+\delta_{\alpha_\beta})N_{\alpha\beta}$ 
(with $\delta_{\alpha_\beta}$ the Kronecker delta) and the total number of sites 
of type $\alpha$ is $2N_\alpha$, then the probability of bonding a site of type
$\alpha$ to one of type $\beta$ is \cite{Tavares:2010}. 
\begin{equation}
\label{eqpalphabeta}
p_{\alpha\beta}=(1+\delta_{\alpha_\beta})\frac{N_{\alpha\beta}}{2N_\alpha}.
\end{equation}
(Notice that, although $N_{\alpha\beta} = N_{\beta\alpha}$ always holds,
if $N_\alpha \neq N_\beta$ then $p_{\alpha\beta} \neq p_{\beta\alpha}$.)
From these probabilities, which can be obtained by solving the laws of mass 
action, equations (\ref{eq:lma01}) and (\ref{eq:lma02}), we can compute a 
number of interesting structural quantities. In particular, we shall derive 
the statistics of `blocks', i.e., the probabilities of assembling sequences 
of contiguous identical bonds (`blocks') of length $\ell_{\alpha\beta}$, defined 
as the number of $\alpha\beta$ bonds in the sequence (block).

Let us consider first blocks of identical particles. To make an $A$ 
block of length $\ell_{AA}$, one starts with a particle of species $A$ 
that has one $A$ site not bonded to another $A$ site: there are $N_A(1-p_{AA})$
such particles (notice that an $A$ site that is not bonded to another $A$ site
could be either unbonded to any site, or bonded to a $B$ site). Then one needs
to make $\ell_{AA}$ bonds, each with probability $p_{AA}$, which gives a factor
of $p_{AA}^{\ell_{AA}}$. Finally, the block ends with an $A$ site that is not 
bonded to any other $A$ site, hence another factor of $(1-p_{AA})$. It follows
that the number of $A$ blocks of length $\ell_{AA}$ is
\begin{equation}
\label{eq:nAA} 
n(\ell_{AA})=N_Ap_{AA}^{\ell_{AA}}\left(1-p_{AA}\right)^2.
\end{equation}
Likewise, the number of $B$ blocks of length $\ell_{BB}$ is
\begin{equation}
\label{eq:nBB} 
n(\ell_{BB})=N_Bp_{BB}^{\ell_{BB}}\left(1-p_{BB}\right)^2.
\end{equation}

Now consider $AB$ blocks of block size $\ell_{AB}$, i.e., alternating sequences 
of $A$ and $B$ particles. Two cases must be distinguished: blocks with either 
$A$ or $B$ sites at both ends have odd $\ell_{AB}$, whereas blocks with an $A$ 
site at one end and a $B$ site at the other end have even  $\ell_{AB}$. 
The number of $AB$ blocks with odd $\ell_{AB}$ is
\begin{equation}
\label{nlABodd}
\begin{split}
n(\ell_{AB},{\rm{odd}}) =&  N_A\left(1-p_{AB}\right)
p_{AB}^{(\ell_{AB}+1)/2}p_{BA}^{(\ell_{AB}-1)/2}\left(1-p_{BA}\right)\\ 
&
+N_B\left(1-p_{BA}\right)p_{BA}^{(\ell_{AB}+1)/2}
p_{AB}^{(\ell_{AB}-1)/2}\left(1-p_{AB}\right).
\end{split}
\end{equation}
The first term on the right-hand side (rhs) of this equation is derived as 
follows: if an $AB$ block starts with an $A$ particle, then one of its $A$ 
sites is not bonded to an $B$ site, which gives the factor 
$N_A\left(1-p_{AB}\right)$; 
then, there follow $\left(\ell_{AB}+1\right)/2$ $AB$ bonds alternating with 
$\left(\ell_{AB}-1\right)/2$ $BA$ bonds, which gives the factor 
$p_{AB}^{(\ell_{AB}+1)/2}p_{BA}^{(\ell_{AB}-1)/2}$; finally, the block ends 
with a $B$ site not connected to an $A$ site, which gives the factor 
$\left(1-p_{BA}\right)$. The second term is obtained by just exchanging $A$ 
and $B$ in the preceding argument: it corresponds to counting the $AB$ and 
$BA$ bonds for an $AB$ block that starts with a $B$ particle and ends with an 
$A$ particle. Equation (\ref{nlABodd}) can be simplified using the definitions 
of $p_{AB}$ and $p_{BA}$, with the result
\begin{equation}
\label{nlabodd2}
n(\ell_{AB},{\rm{odd}}) = 2N_A^{1/2} N_B^{1/2} 
\left(1-p_{AB}\right)\left(1-p_{BA}\right) p_{AB}^{\ell_{AB}/2}p_{BA}^{\ell_{AB}/2}.
\end{equation}

By the same reasoning, the number of $AB$ blocks with even $\ell_{AB}$ is
\begin{equation}
\label{nlABeven}
\begin{split}
n(\ell_{AB},{\rm{even}})= &N_A\left(1-p_{AB}\right)
\left(p_{AB}p_{BA}\right)^{\ell_{AB}/2}\left(1-p_{AB}\right) +\\
&+N_B\left(1-p_{BA}\right)\left(p_{AB}p_{BA}\right)^{\ell_{AB}/2}
\left(1-p_{BA}\right).
\end{split}
\end{equation}
We reiterate that, in order to fulfil symmetry under sequence inversion,
i.e., the requirement that the identity 
of a block should be independent of the order in which its sequence
is read, equations (\ref{nlABodd}) and (\ref{nlABeven}) include a contribution 
from both: $AB\ldots AB$ and $BA\ldots BA$ sequences, for $\ell_{AB}$ odd;
and from $AB\ldots BA$ and $BA\ldots AB$ sequences, for $\ell_{AB}$ even.

The mean block lengths can now be calculated. For $AA$ and $BB$ blocks, we have
\begin{eqnarray}
\label{eq:lAav}
\langle \ell_{AA} \rangle &=&\frac{1}{1-p_{AA}}, \\
\label{eq:lBav}
\langle \ell_{BB} \rangle &=&\frac{1}{1-p_{BB}}.
\end{eqnarray}
Notice that these expressions are general, in the sense that they apply even 
when $N_A\ne N_B$. The mean length of $AB$ blocks is
\begin{eqnarray}
\langle \ell_{AB} \rangle &=& \frac{ \sum_{i=1}^\infty 
\left[(2i-1)n(2i-1,{\rm{odd}})+2i\: n(2i,{\rm{even}})\right]}
{ \sum_{i=1}^\infty \left[n(2i-1,{\rm{odd}})+n(2i,{\rm{even}})\right]}\nonumber\\
\mbox{} &=& \frac{1}{1-p_{AB}p_{BA}}\frac{1+p_{AB}p_{BA}+F(p_{AB},p_{BA})}
{1+\frac{1}{2}F(p_{AB},p_{BA})},
\label{eq:lABav}
\end{eqnarray}
where
\begin{equation}
\label{Fxy}
F(x,y)=\frac{x(1-y)^2+y(1-x)^2}{(1-x)(1-y)}.
\end{equation}
It is readily seen that $\langle \ell_{AA}\rangle $ and $\langle\ell_{BB}\rangle$
are functions of, respectively, $p_{AA}$ and $p_{BB}$ only, whereas  
$\langle\ell_{AB}\rangle$  is a function of $p_{AB}$ and $p_{BA}$ only.
It follows that results do not depend on whether these 
blocks are isolated or part of longer chains. Further note that, by 
construction, the minimum length of an $\alpha\beta$ block is 1, when 
$p_{\alpha\beta}\to 0$: this is because, in this limit, $N_{\alpha\beta} \to 0$
and both the number of $\alpha\beta$ blocks and their length $\to 0$, 
but the ratio of these two quantities $\to 1$. 

If $N_A=N_B$, in which case $p_{AB}=p_{BA}$, equation (\ref{eq:lABav}) 
simplifies to
\begin{equation}
\langle \ell_{AB} \rangle = \frac{1}{1-p_{AB}}.
\end{equation}
Furthermore, equations (\ref{nlABodd}) and (\ref{nlABeven}) become 
identical, leading to
\begin{equation}
n(\ell_{AB})=2N_A\left(1-p_{AB}\right)^2p_{AB}^{\ell_{AB}}.
\end{equation}

\subsection{Extension to multiple species}

The theory of linear aggregation of the preceding section can be 
straightforwardly extended to the case where we have $n$ distinct chemical 
species $\alpha = S_1, S_2, \ldots, S_n$,
each decorated with two identical bonding sites. If, as before, 
$N_\alpha$ is the number of particles of species $\alpha$, the total number
of particles in the system will be $N = \sum_\alpha N_\alpha$, and their mole 
fractions $x_\alpha = N_\alpha /N$. Therefore we have a set of $n(n+1)/2$ coupled 
chemical reactions:
\begin{equation}
	\overline{\alpha} + \overline{\beta} \rightleftharpoons \alpha\beta,
	\qquad \alpha, \beta = S_1,\ldots,S_n,
\end{equation}
subject to the $n$ constraints
\begin{equation}
	2N_\alpha = M_{\overline\alpha} + 2N_{\alpha\alpha} +
	\sum_{\beta\neq\alpha} N_{\alpha\beta},
	\qquad \alpha = S_1,\ldots,S_n.
\end{equation}
In the ideal gas of clusters approximation, the condition of chemical 
equilibrium between any pair of species $\alpha$ and $\beta$ is given by 
(\ref{eq:equilibrium}) and similarly the equilibrium constant of each reaction 
by (\ref{eq:Kab}). The $n$ laws of mass action are obtained as
\begin{equation}
\label{eq:lmagen}
	1-X_\alpha = \sum_{\beta} \left(1+\delta_{\alpha\beta}\right)
	x_\beta X_\alpha X_\beta K_{\alpha\beta},
\end{equation}
where $X_\alpha = M_{\overline\alpha}/2N_\alpha$.
Accordingly, the results in block statistics could be extended to account
for the assembly of more complex architectures.

As mentioned above, we require some prescription for finding the equilibrium
contants $K_{\alpha\beta}$, and thence the probabilities $p_\alpha$ and 
$p_{\alpha\beta}$. For DNA, perhaps the simplest way is to compute them using the
second equality in equation (\ref{eq:Kab}) with $\Delta G_{\alpha\beta}$ given
by Santalucia for the nearest-neighbour model 
\cite{santalucia1998unified,bommarito2000thermodynamic}. Alternatively, one
can map a microscopic, off-lattice theory of self-assembly onto the above
minimal description. This we do in the next section; a similar approach 
has been proposed by Reinhardt and Frenkel \cite{RF}.

\subsection{Wertheim's thermodynamic perturbation theory}

\label{subsect-Wertheim}

Wertheim's thermodynamic perturbation theory (TPT) is a microsopic theory
for the self-assembly of particles interacting via strong, short-ranged 
attractions \cite{Wertheim:1984a,Wertheim:1984b}. It has found novel 
applications in the description of the phase behaviour of patchy colloidal 
particles~\cite{bianchi,russo,starr}. As in \cite{LocRov}, 
we rather bluntly approximate the solution of DNA sequences as a binary 
mixture of $N_A$ and $N_B$ equisized hard spheres (HSs) of diameter $\sigma$, 
contained in a volume $V$; the total number density is thus 
$\rho=(N_A+N_B)/V\equiv N/V$. The solvent is not explicitly considered. 
Both species are divalent: particles of species $A$ are decorated with two 
attractive sites, or `patches', of type $A$, and particles of type $B$ are 
decorated with two patches of type $B$: these represent the single strands 
at the end of the DNA sequences. We make the usual assumption that the patches 
are distributed over the spheres' surfaces in such a way that each patch can 
only take part in at most one bond, which is a short-ranged attractive 
interaction between two patches, as is appropriate for DNA bases. We take 
these inter-patch attractions to be square wells of depth $\epsilon_{\alpha\beta}$
and range chosen such that the volume available to an $\alpha\beta$ bond is 
$v_b^{\alpha\beta}$ ($\alpha,\beta=A,B$).

Following \cite{Heras2} the bonding probabilities $p_{\alpha\beta}$ are given by
\begin{eqnarray}
\label{eq:pAA}
p_{AA}&=&2\xi x_A\Delta_{AA}X_A^2, \\
\label{eq:pBB}
p_{BB}&=&2\xi x_B\Delta_{BB}X_B^2, \\
\label{eq:pAB}
p_{AB}&=&2\xi x_B\Delta_{AB}X_AX_B, \\
\label{eq:pBA}
p_{BA}&=&2\xi x_A\Delta_{AB}X_AX_B,
\end{eqnarray}
where $\xi=\rho v_{HS}$, with $v_{HS}=\pi\sigma^3/6$ the volume of a HS, is the 
(total) packing fraction, $x_\alpha=N_\alpha/N$ is the mole fraction of component 
$\alpha$ ($\alpha=A,B$), and $\Delta_{\alpha\beta}$ are the bond partition 
functions. In the low-density, strong-interaction limit, which as we shall see
is appropriate to our simulations, we have
\begin{equation}
\label{eq:Delta}
\Delta_{\alpha\beta}\approx{v_b^{\alpha\beta}\over v_S}\exp(\beta\epsilon_{\alpha\beta}),
\end{equation}
In equations (\ref{eq:pAA})--(\ref{eq:pBA}), $X_\alpha$, the fractions of 
unbonded sites of type $\alpha=A,B$, are given by the following laws of 
mass action:
\begin{eqnarray}
\label{eq:lma1}
1-X_A&=&2\xi x_B\Delta_{AB}X_AX_B+2\xi x_A\Delta_{AA}X_A^2, \\
\label{eq:lma2}
1-X_B&=&2\xi x_B\Delta_{BB}X_B^2+2\xi x_A\Delta_{AB}X_AX_B.
\end{eqnarray}
Comparing equations (\ref{eq:lma01}) and (\ref{eq:lma1}), (\ref{eq:lma02})
and (\ref{eq:lma2}), and further noting that the fraction of sites of type
$\alpha$ is the same as the mole fraction of particles of species $\alpha$, 
we conclude that the reaction constants in our minimal description are very
simply related to the bond partition functions:
\begin{equation}
\label{eq:Kab4}
K_{\alpha\beta}=\xi\Delta_{\alpha\beta}.
\end{equation}

How can we now relate the parameters of Wertheim's TPT to those of Santalucia?
Start by noting that, from equations (\ref{eq:Kab}),(\ref{eq:Delta}) and 
(\ref{eq:Kab4}), we have
\begin{equation}
\label{eq:DeltaG}
\beta\Delta G_{\alpha\beta}=-\log\left(\xi\Delta_{\alpha\beta}\right)
=-\log\left(\rho v_b^{\alpha\beta} \right)-\beta\epsilon_{\alpha\beta}.
\end{equation}
Recalling that $\Delta G_{\alpha\beta}=\Delta H_{\alpha\beta}-T\Delta S_{\alpha\beta}$,
where $H_{\alpha\beta}$ and $S_{\alpha\beta}$ are, respectively, the enthalpy and 
entropy of an $\alpha\beta$ bond, we can identify
\begin{equation}
\label{eq:deltaH}
\Delta H_{\alpha\beta}=-\epsilon_{\alpha\beta} \qquad , \qquad
\Delta S_{\alpha\beta}=\log\left(\rho v_b^{\alpha\beta} \right), 
\end{equation}
i.e., the change in enthalpy is related to the bond strength, and the change 
in entropy to the volume available to the bond. In actual systems it is often 
the case that the pressure and volume vary very little, and the change in 
enthalpy can thus be equated to a change in internal energy
\cite{Sciortino:2007}.

Wertheim's theory thus provides an inexpensive alternative description of the 
self-assembly statistics in block copolymer systems, on the basis of very 
simple model -- patchy particles --  whose interaction parameters can be 
readily related to hybridisation enthalpies and entropies. It has exactly the 
same structure as the minimal theory of linear aggregation of the preceding 
section, so the same accuracy can be expected.

\section{Model for DNA}

\label{sect-model}

We describe DNA using the oxDNA model \cite{OxDNA1,OxDNA2}. This is a 
coarse-grained model with implicit solvent, which has been shown to capture the 
basic thermodynamics, as well as the essential structural properties, of DNA. 
It consists of rigid nucleotides, interacting via pairwise interactions that 
comprise non-linear elastic, stacking, cross-stacking, excluded-volume and 
hydrogen bonding contributions; see \cite{OxDNA2} for details.

\begin{figure}
\includegraphics[width=\textwidth]{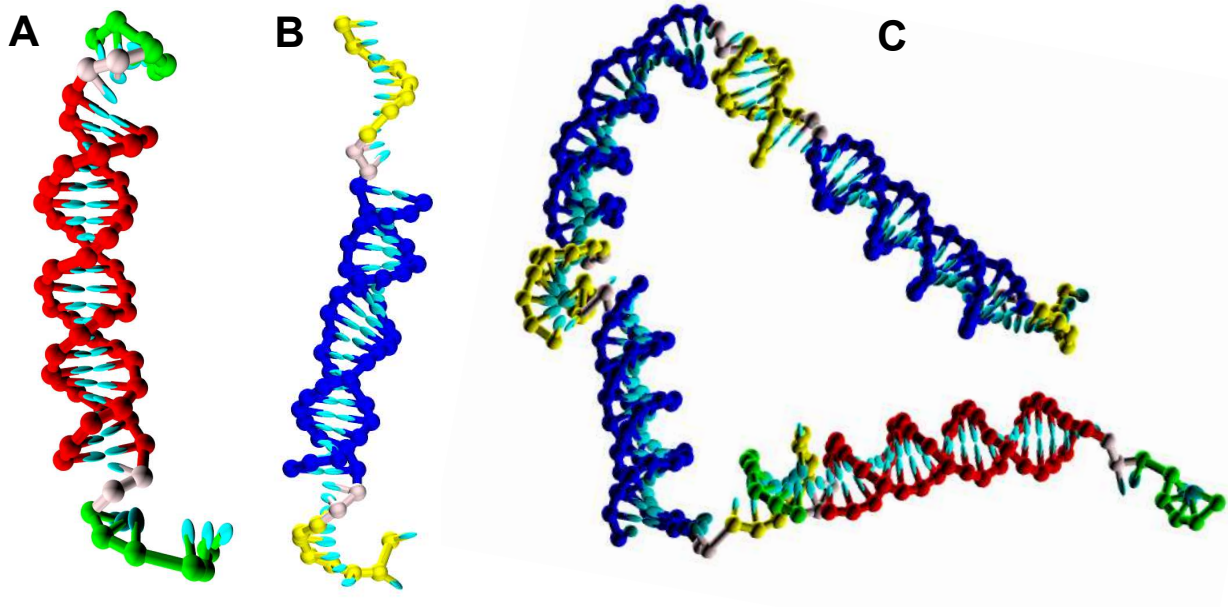}
\caption{Cartoons of particles of (a) species $A$-and (b) species $B$,
made up of a core of double-stranded DNA and two identical
single strands at the ends acting as bonding sites.
On both particles species, two extra bases (adenine, coloured light pink) on
each side of the core have been included to decouple the free ends 
from the core. (c) A chain of four particles, one of species $A$ and three of
species $B$. This $BBBA$ chain is composed of on4 $BB$ block of length 2 and 
one $AB$ block of length 1.}
\label{fig1}
\end{figure}

The system we investigate is a binary mixture 
of DNA nanoparticles~\cite{ns1,ns2}.
Each `particle' is made up of a complementary double helix core 
${\bf X}$ decorated with identical single strands, of types $a$ or $b$, 
at either end, i.e., the two particle species are $A=a{\bf X}a$ and 
$B=b{\bf X}b$. The binding enthalpy of the 
$a\!-\!a$, $b\!-\!b$ and $a\!-\!b$ pairings 
can be tuned via a judicious choice of the sticky-end binding sequences.
In what follows, we shall take $A$ ($B$) to refer interchangeably
to either a single strand of type $a$ ($b$) or a particle of species $A$ ($B$). 
To break the symmetry of the model, we want to favour $BB$ bonds over $AA$ or 
$AB$ bonds. We thus need to find pairs of short self-complementary DNA 
sequences $A$ and $B$ that can bind to each other. Preliminary calculations
using Wertheim's theory suggest rich behaviour is realised if their 
hybridisation enthalpies satisfy the conditions
\begin{equation}
\label{eq:enthalpies}
\frac{\Delta H_{AB}}{\Delta H_{AA}} \simeq 1, \quad 
\frac{\Delta H_{BB}}{\Delta H_{AA}} \simeq 1.25.
\end{equation}
No condition is set on their hybridisation entropies. The nearest-neighbour 
model of SantaLucia \cite{santalucia1998unified} allows us to compute these 
quantities for any given sequence: 
the values of $\Delta H$ and $\Delta S$ can be assumed to be 
temperature-independent at least in a `narrow' range around $T=\SI{310}{\K}$.
In order to fulfil conditions (\ref{eq:enthalpies}), $B$ is chosen to be the 
same as $A$ with an extra pair of complementary nucleotides, one at each end, 
so that $B$ will bind to $B$ with greater energy than $A$ to $A$, but will 
still be able to bind to $A$ with roughly the same energy as $A$ to $A$.
The dangling ends in an $AB$ bond will actually provide an extra contribution, 
usually increasing stability of $AB$ with respect to $AA$, as evaluated in 
\cite{bommarito2000thermodynamic}; without this stabilizing mechanism there 
would be no reason for $A$ and $B$ to form this mixed complex instead of only 
$AA$ and $BB$. This also implies that the condition 
$\frac{\Delta H_{AB}}{\Delta H_{AA}} \simeq 1$ is not really attainable, so we 
should just look for this ratio to be as close as possible to unity. Two 
sequences that come close to these target values are $A$=CGATCG and 
$B$=TCGATCGA, whose hybridisation enthalpies and entropies are reported in 
Table~\ref{table1}. Note that $A$ can bind to $A$ 
(being `palindromic') with six bases, $B$ can bind to $B$ with eight bases, 
and $A$ can bind to $B$ with six bases. A cartoon of a particle is shown in 
figure \ref{fig1}.

\section{Results}

\label{sect-results}

We ran constant-volume molecular dynamics (MD) simulations starting with 
$N_A=N_B=200$ bifunctional particles of types $a{\bf X}a$ and $b{\bf X}b$, 
for a total of 800 sticky ends on 400 particles. The resulting concentration 
of single strands of each species in solution is $c \approx 392 \,\mu{\rm M}$.
We mimicked (recall there is no explicit solvent) an 
aqueous solution of salt concentration $[\ce{Na+}]=\SI{0.4}{\molar}$, at 
temperatures in the range $[40,58]\si{\celsius}$, where binding is expected to 
take place and equilibration can be achieved with the available computational 
resources.

From the simulation, we estimate the number of $A$ patches unbonded, bonded 
with $A$ ($N_{AA}$) and bonded with $B$ ($N_{AB}$). Similar quantities are 
calculated for $B$ patches. Using the relations previously introduced we 
can then estimate $p_{AA} \equiv N_{AA}/N_A$, $p_{BB} \equiv N_{BB}/N_B$ and
$p_{AB} \equiv \frac{N_{AB}}{N_A}$, as well as $X_{A}$ and $X_{B}$.

\begin{figure}
\includegraphics[width=\textwidth]{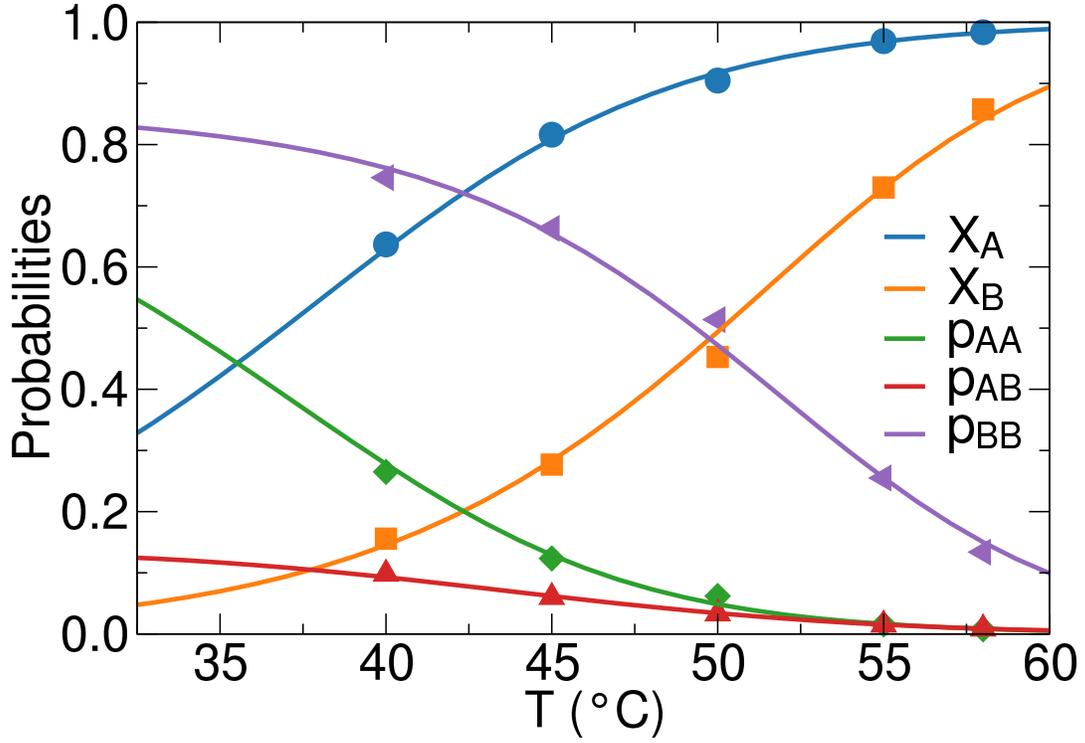}
\caption{Probabilities $p_{AA}, p_{AB}=p_{BA}$ and $p_{BB}$ of the different types
of bonds between patches, and probability $X_A$ ($X_B)$ that a $A$ ($B$) site 
is unbonded, {\it vs} temperature, from simulations (symbols) and theory 
(solid lines). The solid lines were obtained by solving equations 
(\ref{boundfractions}) numerically using $\Delta H_{ij}$ and $\Delta S_{ij}$ in 
table \ref{table2} } 
\label{fig2}
\end{figure}

The resulting probabilities are shown in figure \ref{fig2} as symbols; 
the curves are the theory predictions computed 
as explained below. The most difficult part of the 
simulation is to `equilibrate' the $BB$ bonds, which, being composed
of more bases, are stronger and rarely break. The next-stronger bonds are $AB$ 
and $AA$, respectively. However, since forming $AB$ bonds would decrease the 
number of $BB$ bonds, the system as a whole prefers to form more $AA$ bonds 
than $AB$ bonds, in order to free $B$ patches for the most favourable $BB$ 
bonds. 

\begin{figure}
\includegraphics[width=\textwidth]{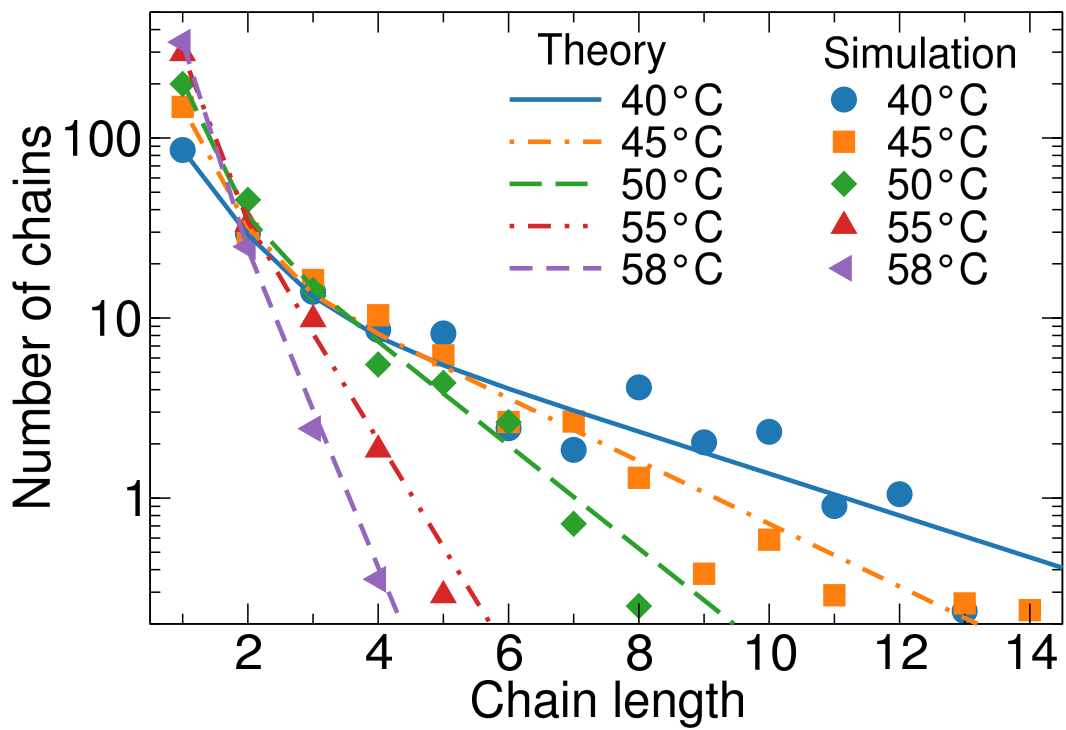}
\caption{Comparison between expected and simulated chain length
distributions at all temperatures studied.}
\label{fig3}
\end{figure}

From $ p_{AA}$, $p_{BB}$ and $p_{AB}$ we can then predict the chain length 
distribution $P(n)$, where $n$ is the number of particles in the chain.
Let $s$ be a sequence of bonded monomers ($A$ or $B$ particles); the 
probability of observing that sequence is
\begin{equation}
\label{prob}
	P(s) = \qty( 1-p_A )^{m_{\overline{A}}}\qty( 1-p_B )^{m_{\overline{B}}} 
p_{AB}^{n_{AB}} p_{BA}^{n_{BA}} p_{AA}^{n_{AA}} p_{BB}^{n_{BB}} ,
\end{equation}
where $m_{\overline\alpha}$ is the number of free ends of type $\alpha$ and 
$n_{\alpha\beta}$ is the number of sites of type $\alpha$ bonded to sites of 
type $\beta$ (with the constraints $m_{\overline{A}}+m_{\overline{B}}=2$,
$n_{AA} + n_{AB} +  n_{BB}= n-1$). Then the probability $P(n)$ of observing a 
cluster (literally a linear chain) of length $n$) is found by summing $P(s)$ 
over all possible sequences of $n$ monomers: 
\begin{equation}
\label{eqPofn}
P(n)=\sum_{s=1}^{2^n} P(s).
\end{equation}
Figure \ref{fig3} compares theoretical predictions and simulation data for the 
chain length distribution at all temperatures studied.

\begin{figure}
\includegraphics[width=\textwidth]{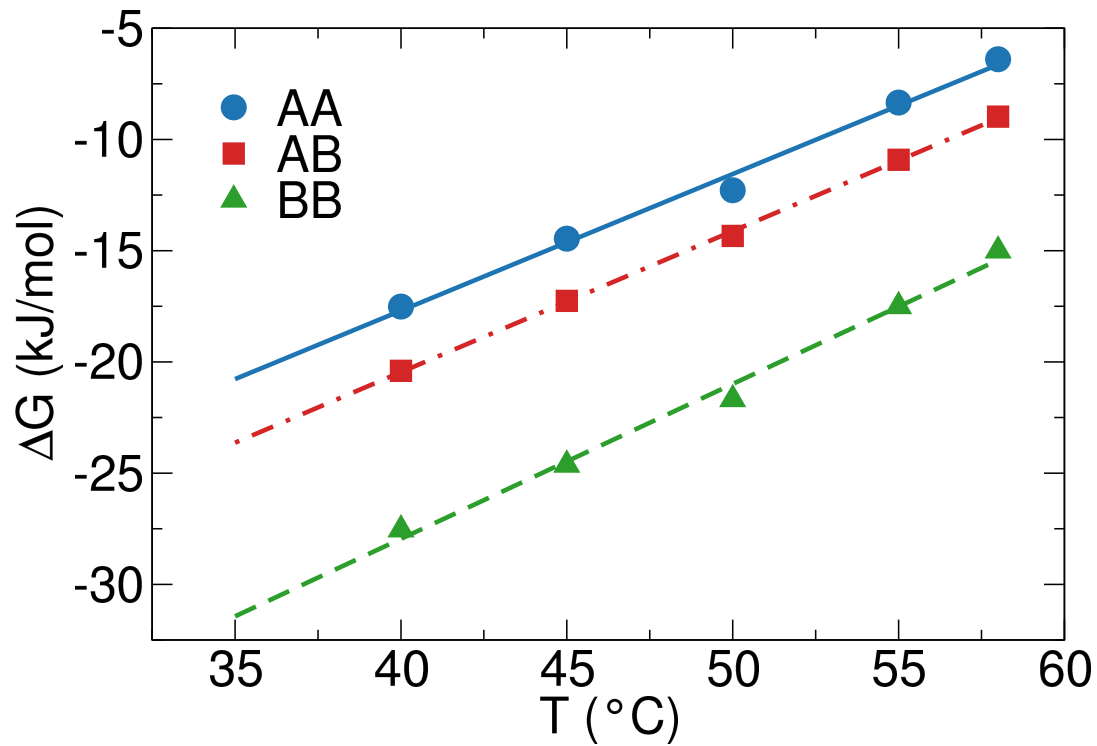}
\caption{Binding free energy of the three complexes $AA$, $BB$ and $AB$. 
Symbols are the $\Delta G_{\alpha\beta}$ values obtained by solving equation
(\ref{boundfractions}) using the $p_{\alpha,\beta}$ from simulation.
Solid lines are the linear fits $\Delta G_{\alpha\beta} = 
\Delta H_{\alpha\beta}- T\Delta S_{\alpha\beta}$, where $\Delta H_{\alpha\beta}$ and
$\Delta S_{\alpha\beta}$ are taken as temperature-independent fit parameters 
(values reported in table \ref{table2}).}
\label{fig5}
\end{figure}

To evaluate $\Delta G_{\alpha\beta}$  we first combine equations (\ref{eq:Kab}),
(\ref{eq:pAA})--(\ref{eq:pAB}) and (\ref{eq:Kab4}) to obtain, for $x_A=x_B=0.5$,
\begin{equation}
\label{boundfractions}
p_{\alpha\beta}=X_\alpha X_\beta\exp\left(-\beta\Delta G_{\alpha\beta}\right).
\end{equation}
Equations (\ref{boundfractions}) can now be solved
using the data in figure \ref{fig2}, at each $T$. Results are plotted
in figure \ref{fig5}. In all cases, a linear dependence of 
$\Delta G_{\alpha\beta}$ on $T$ is observed. The slope and intercept provide 
the best-fit values for $\Delta H_{\alpha\beta}$ and $\Delta S_{\alpha\beta}$,
which can then be used backwards to predict the bond probabilities. These 
predictions are shown as solid lines in figure \ref{fig2}.

Table \ref{table2} compares the best-fit values of $\Delta H_{\alpha\beta}$ and 
$\Delta S_{\alpha\beta}$ with the predictions of the SantaLucia model. Though not
excellent, agreement is reasonable, considering that the oxDNA model is a 
parametrisation based on SantaLucia estimates for the melting temperature. 
Specifically, oxDNA predictions for the melting temperatures have been found 
to deviate on average \SI{1.4}{\celsius} from those of SantaLucia
\cite{Sulc:2012}.
  
\begin{figure}
\includegraphics[width=\textwidth]{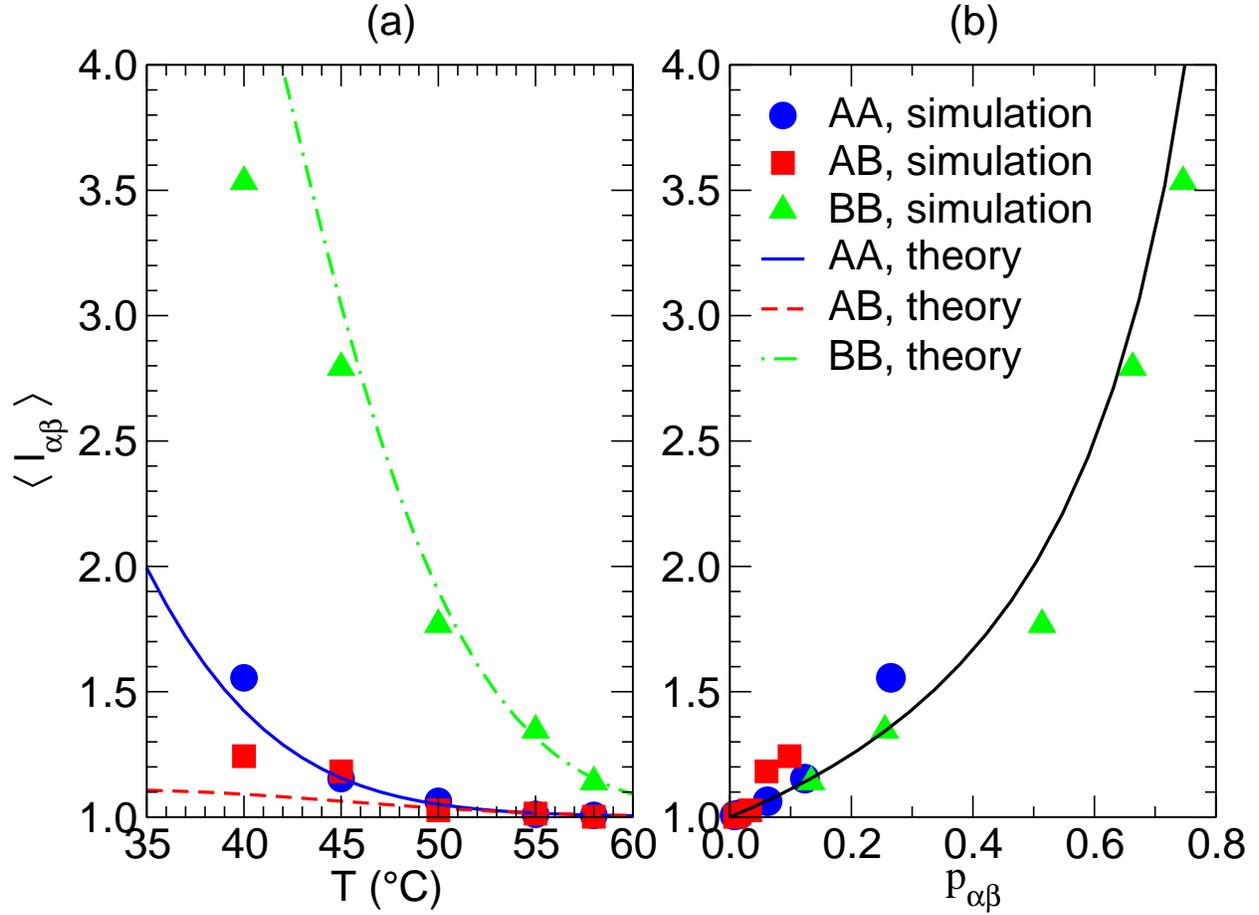}
\caption{Mean block lengths $\langle \ell_{\alpha\beta}\rangle$ {\it vs}
(a) temperature; (b) bond probability $p_{\alpha\beta}$. In (b) the black 
solid line is the theory prediction $1/(1-p_{\alpha\beta})$  for all block 
types, e.g., equations (\ref{eq:lAav})--(\ref{eq:lABav}) when $p_{AB}=p_{BA}$ }
\label{fig6}
\end{figure}

Finally, figure \ref{fig6} plots the mean block lengths 
$\langle \ell_{\alpha\beta}\rangle$, {\it vs} either temperature 
(figure \ref{fig6}a) or bond probability $p_{\alpha\beta}$ (figure \ref{fig6}b).
For our choice of parameters, both $AA$ and $AB$ blocks are very short and 
the simulation data are very noisy; agreement between theory and simulation is
encouraging for the longer $BB$ blocks.

\section{Conclusions}

\label{sect-conclusions}

We have proposed a minimal theoretical framework for the assembly of linear 
block copolymers. This makes very few assumptions on the nature of the 
monomers, namely: (i) assembly is assimilated to reversible chemical reactions 
between short-ranged bonding sites; (ii) each site can participate in at most 
one bond; and (iii) the overall concentration is low enough that sites and 
bonds behave as ideal gases. The theory requires as inputs the reaction 
constants for the polymerisation reactions. Importantly, these can be 
derived from theories that incorporate only very limited detail of the 
actual molecular processes.

The theory was tested against simulation results for the assembly of DNA chains 
from two types of short duplexes, as described by the oxDNA model, using 
reaction constants calculated from Santalucia's theory of a lattice model 
of DNA. This was found to reproduce the equilibrium block size distributions,
mean block sizes, and fractions of unreacted monomers fairly well. 

The theory is easily generalised to any number of associating particle species
in any proportion. In our view it has the potential to become a useful tool
to predict or reverse-engineer the architectures of multi-block copolymers or
polycolloids, and even provide some insight into the kinetics of association.

\section*{Conflicts of interest}

There are no conflicts to declare.

\begin{acknowledgments}

R. F. and F. S. acknowledge support from MIUR PRIN 2017 (Project 2017Z55KCW).
J. M. T. and P. I. C. T. acknowledge financial support from the Portuguese 
Foundation for Science and Technology (FCT) under Contracts nos.\ 
UIDB/00618/2020 and UIDP/00618/2020. 

\end{acknowledgments}

\newpage

\begin{table}[htp]
\centering
\begin{tabular}{|c|c|c|c|c|c|}
\hline
$A$  &$B$  &$\frac{\Delta H_{BB}}{\Delta H_{AA}}$  &$\frac{\Delta H_{AB}}{\Delta H_{AA}}$  &$\frac{\Delta S_{BB}}{\Delta S_{AA}}$  &$\frac{\Delta S_{AB}}{\Delta S_{AA}}$ \\
\hline
CGATCG   &TCGATCGA   &1.27   &1.09   &1.24   &1.09 \\
\hline
\end{tabular}
\vspace{0.5cm}
\caption{Relative hybridisation enthalpies and entropies of DNA single strands
used in our simulations, calculated according to \cite{santalucia1998unified} 
and \cite{bommarito2000thermodynamic}.}
\label{table1}
\end{table}

\newpage


\begin{table}[htp]
\centering
\begin{tabular}{|c|c|c|c|}
\hline
{}                        &$AA$          &$BB$          &$AB$ \\
\hline
SantaLucia             &              &              &         \\
$\Delta S$ (J/mol/K)      &$-126.2$      &$-156.8$      &$-137.1$ \\
$\Delta H$ (kJ/mol)       &$-44.6$       &$-56.6$       &$-48.6$  \\
\hline
Simulation             &              &              &         \\
$\Delta S$ (J/mol/K)      &$-146.8$      &$-166.4$      &$-151.5$ \\
$\Delta H$ (kJ/mol)       &$-50.2$       &$-58.8$       &$-52.3$  \\
\hline
\end{tabular}
\vspace{0.5cm}
\caption{Comparison of best-fit values of $\Delta H_{\alpha\beta}$ and 
$\Delta S_{\alpha\beta}$ with the predictions of the SantaLucia model.}
\label{table2}
\end{table}

\newpage

\newpage
\newpage
\newpage
\newpage

\end{document}